\documentstyle[aps,prl,epsf]{revtex}

\begin{document}
\draft



\twocolumn[\hsize\textwidth\columnwidth\hsize\csname %
@twocolumnfalse\endcsname
%


\title{Proton NMR for Measuring Quantum-Level Crossing\\ in the Magnetic Molecular Ring Fe10}


\author{M.-H. Julien$^{1,2}$, Z.H. Jang$^1$, A. Lascialfari$^2$, 
F. Borsa$^{1,2}$, M. Horvati\'c$^3$, A. Caneschi$^4$, and D. Gatteschi$^4$}


\address{$^1$Department of Physics and Astronomy, Ames Laboratory,
Iowa State University, Ames, Iowa 50011}
\address{$^2$Dipartimento di Fisica "A. Volta" e Unit\'a INFM di Pavia,
Via Bassi 6, 27100 Pavia, Italy}
\address{$^3$Grenoble High Magnetic Field Laboratory, CNRS and MPI-FKF,
BP 166, 38042 Grenoble Cedex 9, France}
\address{$^4$Department of Chemistry, University of Firenze, Via Maragliano 77,
50144 Firenze, Italy}


\date{March 9, 1999}
\maketitle


\widetext


\begin{abstract}


The proton nuclear spin-lattice relaxation rate 1/$T_1$ has been measured as
a function of temperature and magnetic field (up to 15 T) in the 
molecular magnetic ring Fe$_{10}$(OCH$_{3}$)$_{20}$(O$_{2}$
CCH$_{2}$Cl)$_{10}$ (Fe10). Striking enhancement of 
1/$T_1$ is observed around magnetic field values corresponding to
a crossing between the ground state and the excited states
of the molecule.
We propose that this is due to a cross-relaxation 
effect between the nuclear Zeeman reservoir
and the reservoir of the Zeeman levels of the 
molecule. This effect provides a powerful tool to investigate 
quantum dynamical phenomena at level crossing.


\end{abstract}


]
\narrowtext



The magnetic properties of metal ion clusters incorporated in
large molecules 
attract considerable interest for the new physics involved
and for the potential 
applications \cite{Gatteschi,Awschalom95}. At low temperatures, these molecules act as
individual quantum nanomagnets, enabling
to probe, at the macroscopic scale, the crossover between quantum and classical physics
\cite{Stamp96}.
Of fundamental interest is the situation of (near-) degeneracy of two magnetic levels,
where quantum mechanical phenomena such as tunneling or coherence can occur.
These effects have been intensively explored in the recent years, mostly in the
high-spin ($S$=10) molecules Mn12 and Fe8 \cite{Mn12Fe8}, or in the ferritin protein
\cite{Gider95}.
Another interesting system is the molecule
[Fe$_{10}$(OCH$_{3}$)$_{20}$(O$_{2}$CCH$_{2}$Cl)$_{10}$] (in short Fe10), where
the ten Fe$^{3+}$ ions ($s$=5/2) are coupled in a ring 
configuration by an antiferromagnetic (AF) exchange $J/k_B\simeq$13.8~K \cite{Taft94,Zeng97}.
Unlike Mn12 or Fe8, the ground state of Fe10 is nonmagnetic (total spin $S$=0).
The energies $E$ of the excited states are given approximately by Land\'e's rule:
\begin{equation}
E(S) = \frac{P}{2}~S(S+1)
\end{equation}
where $S$ is the total spin value and $P$=$4J/N$, with $N$=10 the number of magnetic ions in the 
ring. In zero magnetic field, the first excited state is $S$=1,
the second $S$=2, etc. (see Fig 1.).
This picture is modified by an external magnetic field,
which lifts the degeneracy of the magnetic states.
A sufficiently strong field can induce level crossings between
the ground state and the excited states, as shown in Fig. 1.
In other words, the ground state of the molecule
can be changed by the field,
from $S$=$0$ to $S$=$1$, then from $S$=$1$ to $S$=$2$, etc. 
Owing to the relatively low value of the magnetic exchange
coupling in Fe10, this field-induced transitions can be observed
experimentally in conventional magnetic fields, for instance through
steps of the magnetization \cite{Taft94,Cornia99}.


The situation of degeneracy between levels raises fundamental problems 
of quantum dynamics \cite{Stamp98,Miyashita98} (specific calculations for
Fe10 can be found in \cite{Chiolero98}). A crucial issue is the role played by the coupling
between magnetic molecular levels and the environment such as phonons and/or nuclear spins
\cite{Stamp98}. Clearly, essential information on this problem
should be accessed through measurements of the nuclear spin-lattice relaxation rate 1/$T_1$
since the nuclei (here protons)
probe the fluctuations of the local field induced at the nuclear
site by the localized magnetic moments.


The physics of level crossings is almost not documented experimentally,
due to the rarity of systems in which the observation is possible.
A situation which has some analogy with the one reported here is the
crossover from antiferromagnetic to ferromagnetic phase in 1D chains,
where a divergence of the one-magnon density of states generates
an enhancement in the nuclear spin-lattice relaxation rate \cite{Azevedo80}.
A closer situation of level crossing between singlet and triplet states can 
be observed in 1D gapped quantum magnets \cite{Chabous}, but the physical context and the
continuum of excited states makes the situation certainly not comparable to that in
finite-size magnets.
In this respect, the mesoscopic ring Fe10 constitutes a model system since magnetic
levels are sharp and well-defined in energy, due to the finite size of the system.


Previous $^1$H NMR relaxation measurements in Fe10 have concerned
magnetic fields much lower than the expected energy gap
$E(1)$$\sim$6~K (Eqn. 1) \cite{Ale97,Ale98a}.         


Here, we present new proton $T_1$ measurements in Fe10,
as a function of magnetic field up to 15 Tesla,
and in the temperature range 1.3~K$\leq$$T$$\leq$4.2~K.
Our main result is the observation of a dramatic enhancement
of $1/T_1$ when the magnetic field reaches the critical values
for which the magnetic levels become degenerate (level-crossing) \cite{rem}.
Although broadening effects due to the use of a powder 
sample prevent yet a quantitative interpretation of the data,
it is pointed out that the cross-relaxation effect between
(proton) nuclear and molecular levels, discovered 
here, should provide a powerful method to investigate the physics of
level-crossing if large enough single crystals become available.


The powder samples were synthesized as described elsewhere \cite{Taft94}.
High-field ($H$$\geq$8~T) NMR measurements were performed
at the Grenoble High Magnetic
Field Laboratory in a 17 T variable field superconducting magnet. 
All measurements where performed with home-built pulsed NMR spectrometers.


\vspace{-9mm}
\begin{figure}[t!]
\begin{center}
\epsfxsize=80mm
 $$\epsffile{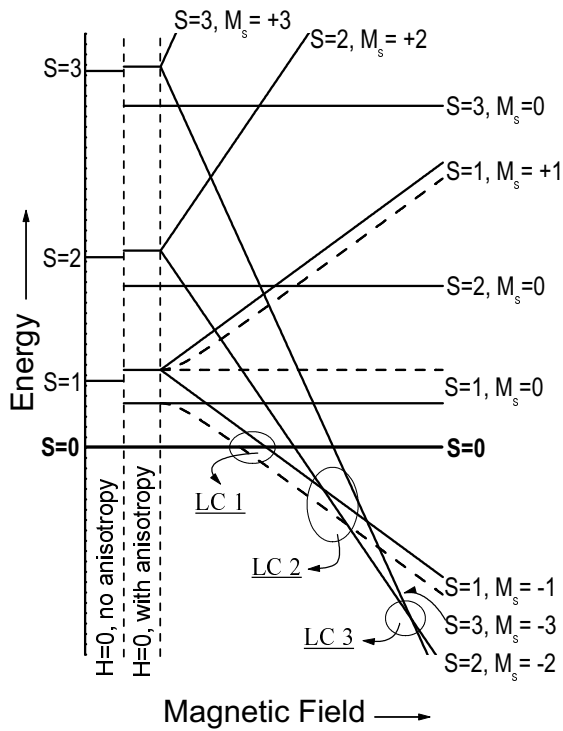}$$
\caption{Energy levels vs. magnetic field for the lower four manifolds
($S$=0 to $S$=3) in Fe10.
The zero-field splitting due to magnetic anisotropy is included only for
the levels relevant to level crossing effects.
Dashed lines are energy levels for $\theta$=90$^{\rm o}$
($S$=1 case). All the other energy levels are for
$\theta$=0$^{\rm o}$. Note that the labels of magnetic levels refer
to the full lines only.
LC1, LC2 and LC3 refer to the three level-crossings
evidenced in this work through proton spin-lattice relaxation.}
\label{levels}
\end{center}                             
\end{figure}


The proton NMR spectrum is featureless, except for an asymmetry related
to the orientation distribution of the grains and to the
superposition of resonaces from inequivalent proton sites in each molecule.
The width of the spectrum is both 
temperature and field dependent due to an inhomogeneous component, {\it i.e.} 
a distribution of hyperfine (dipolar) fields from Fe moments \cite{Ale97}.
At low field ($H$=0.33~T), the full width at 
half maximum (FWHM) is about 25~kHz at room temperature,
it increases to a maximum of about 70~kHz at about 30~K
and it decreases again at low temperature reflecting the collapse
of the spin suceptibility when the Fe10 molecular states condense into the
$S$=0 ground state.
In the temperature range investigated here (1.3~K-4.2~K),
there is a residual field-dependent 
inhomogeneous broadening of the proton NMR line,
which is due to the Fe moments in the $S$=1 excited state.
At 1.3~K the FWHM varies from 25~kHz at $H$=0.33~T to 1.8~MHz 
at 14.65~T. 


$T_1$ was extracted from the recovery 
of the spin-echo amplitude following a sequence of saturating
radiofrequency pulses.
Both $\left(\frac{\pi}{2}\right)_x$-$\left(\frac{\pi}{2}\right)_y$
(solid echo)
and $\left(\frac{\pi}{2}\right)_x$-$\left(\pi\right)_y$
(Hahn echo) sequences were used with similar results.
The recovery of the nuclear magnetization was found to be
non-exponential at all fields. 
For low fields ($H$$\leq$1~T), the NMR line is sufficiently narrow to be 
completely saturated by the radio frequency pulses.
In this case, the non-exponential 
recovery is solely related to the distribution of relaxation rates,
due to the superposition of inequivalent proton sites, and to the
orientation distribution in the powder.
At higher fields, the line becomes
too broad to be completely saturated and thus the initial
recovery is affected by spectral diffusion effects. 
Therefore, in order to measure a relaxation parameter consistently we
chose to define $T_1$ as 
the time at which the nuclear magnetization has recovered half of the
equilibrium value, after removal of the initial fast recovery due
to spectral diffusion. This criterion is 
insensitive to the spectral diffusion, the strength of which depends
on hardly controllable experimental parameters.
The criterion also makes the $T_1$ value insensitive 
to slight modifications of the recovery law that were sometimes observed
for the very long time delays. Otherwise, the shape of the recovery law
was found to be field and $T$-independent.
$T_1$ was also checked to be the same at different positions on the line. 



The magnetic field dependence of proton 1/$T_1$ is reported in Fig.~2.
For technical reasons, experiments between 8 and 15 Tesla were performed at
$T$=1.3~K, while those at lower fields were at $T$=1.5-1.7~K.
The difference is minor and, as will be
seen later, $T_1$ is basically $T$-independent in most of the field range.
So, Fig.~2 can be regarded as the field dependence of $T_1$ at
fixed temperature.
$1/T_1$ shows three very well-defined peaks centered
around the critical field values: 4.7~T, 9.6~T and 14~T.
These values correspond very closely to the fields for 
which steps were observed in the magnetization \cite{Taft94,Cornia99}.


At low fields ($H$$<$1.5~T), the
$T$-dependence of 1/$T_1$ is almost exponential (Fig.~3).
This implies that the proton relaxation is dominated by the singlet-triplet gap and the
finite lifetime of the $S$=1 excited state which generates fluctuations
in the local hyperfine field at the proton site \cite{Chabous,Ale98c}.
The exponential $T$-dependence is a consequence
of the Boltzmann distribution of the $S$=1 population.


However, as shown in Fig.~3, 1/$T_1$ at higher magnetic field appears
to be temperature independent both
at level crossings (4.7 and 9.61~T) and in-between them (7.96~T).
Thus, the strong enhancement around level crossing requires a
new description of the nuclear relaxation, which cannot be based on
thermal excitations.
\vspace{-10mm}
\begin{figure}[t!]
\begin{center}
\epsfxsize=85mm
 $$\epsffile{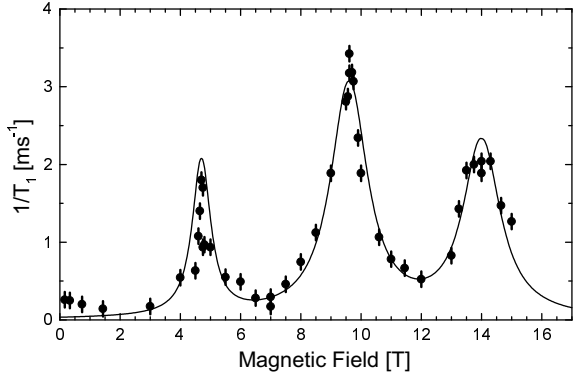}$$
\caption{Magnetic field dependence of proton 1/$T_1$ at 1.3-1.7~K.
The line is a theoretical fit 
according to Eqn. 4 with choice of parameters discussed in the text.}
\label{champ}
\end{center}
\end{figure}
\vspace{-5mm}
\begin{figure}[t!]
\begin{center}
\epsfxsize=95mm
 $$\epsffile{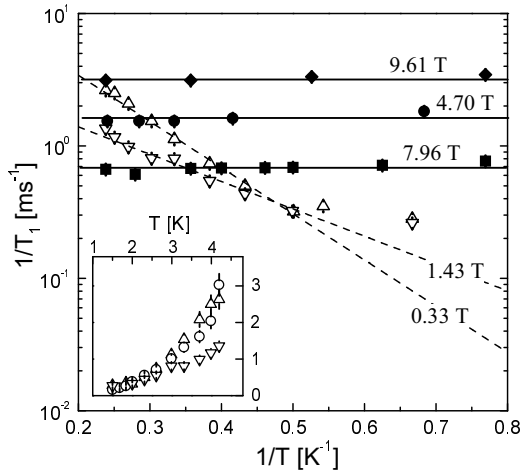}$$
\caption{Main panel: Proton 1/$T_1$ vs. inverse temperature;
activated behaviour at low fields : 0.33~T and 1.43~T, and
constant at 4.7~T, 7.96~T and 9.61~T.
Inset: activated behaviour of 1/$T_1$ in linear scales;
0.33~T ($\triangle$), 0.75~T ($\circ$) and  1.43~T ($\nabla$).}
\label{temperature}
\end{center}
\end{figure}
Near the critical field for level-crossing, the coupled system
nuclei plus molecular magnetic moments can undergo flip-flop energy conserving
transitions, resulting in a transfer of energy from the nuclear system
to the molecular magnet which depends on the matching of energy levels
and not on temperature.
Thus, we propose that the peaks in 1/$T_1$ {\it vs.} magnetic field
are the result of a cross-relaxation effect between the nuclear Zeeman
levels and the magnetic molecular levels.
In fact, since the magnetic molecules are strongly coupled to the "lattice",
the cross-relaxation becomes a very effective channel for spin-lattice relaxation.
It is emphasized that cross-relaxation, here in the sense of matching of energy levels,
is observed between two nuclear reservoirs \cite{Slichter}
or between two electron reservoirs \cite{Abragam}.
Strictly speaking the cross-relaxation occurs only when the condition 
$\hbar\omega_n$=$\hbar\gamma_nH$=$g\mu_B|H-H_c|$ is met.
However, the broadening of both the NMR line 
and of the molecular energy levels can allow the energy conserving
condition to be met over a wide field interval. Furthermore, 
broadening effects are expected for a powder sample.


In order to analyze the data quantitatively, it is
necessary to have a precise description of the magnetic
level diagram for Fe10.
For the triplet state, 
the energy levels are obtained from the diagonalization of the Hamiltonian:
\begin{equation}
{\cal H} = \vec{S} \:\widehat{D}\: \vec{S} +g\mu_B \vec{B}\cdot\vec{S} +{P} ,
\end{equation}


which yields secular equation for energy $E$:
\begin{eqnarray}
(P-\frac{2}{3}D_1-E)(P+\frac{1}{3}D_1-E)^2 \nonumber\\
-(P-\frac{2}{3}D_1-E)g^2\mu_B^2B^2\cos^2\theta \nonumber\\
-(P+\frac{1}{3}D_1-E)g^2\mu_B^2B^2\sin^2\theta = 0 ,
\end{eqnarray}
where we have assumed a diagonal, traceless, axial tensor
for the zero-field splitting (-1/3$D_1$, -1/3$D_1$ , 2/3$D_1$). 
The axis perpendicular to the Fe10 ring plane is a hard axis,
{\it i.e.} $D_1$$>$0.
The values $P$=6.5~K and $D_1$=3.23~K are obtained
from recent torque magnetometry measurements \cite{Cornia99}. 
As shown in Fig.~1,
the critical field $H_c$ for the first level-crossing
depends on the angle $\theta$ between the crystal
field axis and the magnetic field:
$H_c$ varies from 4.33~T for $\theta$=90$^{\rm o}$ up to
5.6~T for $\theta$=0$^{\rm o}$ \cite{Cornia99}. 
This implies a powder distribution of relaxation rates which should
contribute to the width of the first peak at 4.7~Tesla.
The calculation of the level distribution for $S$$\geq$2 is more complex,
making a quantitative analysis of the second and third crossings beyond
the scope of the present paper.


It is very interesting to point out the differences between the
three peaks in 1/$T_1$. At the first level crossing ($H_c$=4.7~T),
there is a very steep increase of 1/$T_1$ occuring in an extremely
narrow field interval (about 0.1~Tesla). This is very suggestive
of a resonant process in the relaxation. The two other peak have a more
regular shape but the third peak is smaller than
the second one.
Of course, we speculate that these differences are 
related to the different spin values involved in each level-crossing.
In particular, the first crossing involves the non-magnetic
level $S$=0.


We tentatively describe the results in Fig. 2 as a sum of Lorenzian 
functions of width $\Gamma_\alpha$ with $\alpha$=1,2,3
for the three level-crossing conditions:
\begin{equation}
\frac{1}{T_1} \propto \sum_{\alpha=1}^{3} A_\alpha 
\left[\frac{\Gamma_\alpha}{\Gamma^2_\alpha+(\gamma_nH-\frac{1}{\hbar}g\mu_B
|H-H_c|)^2}\right]
\end{equation}
This expression fits the data reasonably well with choice of parameters:
$A_1$$\simeq$0.3$A_2$$\simeq$0.5$A_3$=4$\pi$10$^{13}$ rad~s$^{-2}$ ($\equiv$0.36~T)
and $\Gamma_1$$\simeq$0.5$\Gamma_2$$\simeq$0.5$\Gamma_3$=2$\pi$10$^{10}$ rad~s$^{-1}$,
and critical fields $H_{c1}$=4.7~T, $H_{c2}$=9.6~T, $H_{c3}$=14.0~T.
The physical meaning of the coupling constant 
$A_\alpha$ is not clear without a quantitative theory
for the cross-relaxation effect.
The width of each peak is most likely related to
the distribution of level crossing fields
due to the distribution of angles between the magnetic
field and the crystalline axis in our powder sample.
Thermal broadening is also expected since 1.3~K is equivalent
to $\sim$1~T.


In summary we have presented an investigation of the proton spin-lattice
relaxation rate 1/$T_1$ at low 
temperature in the Fe10 molecular magnetic ring.
1/$T_1$ at low fields is dominated by the
thermal fluctuations in the triplet excited state.
At high magnetic fields we have reported a dramatic enhancement of 
the 1/$T_1$ in correspondence to the critical fields for which
the lowest lying molecular energy levels become almost degenerate.
The effect can be explained by a $T$-independent resonant cross-relaxation
effect where thermal fluctuations mediated by phonons do not seeem to
play a role. Thus, the magnetic transitions between nearly degenerate
$\Delta$$S$=1 states become possible, presumably
because of the coupling with the nuclear spins \cite{Rettori,Kiefl86}.


The most promising perspective open by these results concern the
possibility to study dynamical effects of quantum mechanical origin, that are expected
in the vicinity of the level crossing conditions.
Enhanced transfer of population
between two levels is possible, through a mechanism
of quantum tunneling. We have shown here that the dynamics of
nearly degenerate molecular levels is coupled to the dynamics
of nuclear spins. This has to be taken into account in future theoretical
works on Fe10, and at the same time the coupling between nuclei and molecular levels
makes such NMR experiments a privileged tool for detailed studies
when large enough single crystals become available.



Thanks are due to E. Lee for very helpful assistance,
and to A. Rettori, C. Berthier and 
I. Svare for discussions and suggestions. This work has been
partially supported by the "Molecular Magnet" program of the
European Science Foundation, and by the 3MD EU Network (contract
No. ERB 4061 PL-97-0197).
Ames Laboratory is operated for U.S Department of Energy by Iowa
State University under Contract No. W-7405-Eng-82. The work at 
Ames Laboratory was supported by the director for Energy Research,
Office of Basic Energy Sciences. The GHMFL is
Laboratoire Conventionn\'e aux Universit\'es J. Fourier et INPG
Grenoble~I.


\vspace{-0.5cm}






\begin {references}
\bibitem{rem} Throughout this paper, we shall use the term "level-crossing".
Actually, levels may cross or anti-cross (level repulsion) depending on
details of the magnetic Hamiltonian (see for instance Ref. \cite{Kiefl86}).
\end{references}
\end{document}